# Causal inference of post-transcriptional regulation timelines from long-read sequencing in Arabidopsis thaliana

**Rubén Martos**
LaMME
Université d'Évry Paris-Saclay
Évry (France)
ruben.martosprieto@univ-evry.fr

**Christophe Ambroise**
LaMME
Université d'Évry Paris-Saclay
Évry (France)
christophe.ambroise@univ-evry.fr

**Guillem Rigaill**
IPS2, LaMME
Université Paris-Saclay, CNRS, INRAE, Université d'Évry
Orsay (France)
guillem.rigaill@inrae.fr

## ABSTRACT

We propose a novel framework for reconstructing the chronology of genetic regulation using causal inference based on Pearl's theory. The approach proceeds in three main stages: causal discovery, causal inference, and chronology construction. We apply it to the *ndhB* and *ndhD* genes of the chloroplast in *Arabidopsis thaliana*, generating four alternative maturation timeline models per gene, each derived from a different causal discovery algorithm (HC, PC, LiNGAM, or NOTEARS). Two methodological challenges are addressed: the presence of missing data, handled via an EM algorithm that jointly imputes missing values and estimates the Bayesian network, and the selection of the $\ell_1$-regularization parameter in NOTEARS, for which we introduce a stability selection strategy. The resulting causal models consistently outperform reference chronologies in terms of both reliability and model fit. Moreover, by combining causal reasoning with domain expertise, the framework enables the formulation of testable hypotheses and the design of targeted experimental interventions grounded in theoretical predictions.

...

## Table of contents

# 1 Introduction

Transcriptomic regulation in chloroplasts involves a cascade of maturation events—transcript initiation, processing, splicing, trimming, and decay—that unfold over time. Long-read sequencing data now make it possible to observe these events in full transcripts, offering a unique opportunity to study their dependencies and to reconstruct their chronological and possibly causal relationships [1].

Our main contribution in this paper is to provide a **causal inference framework** to order post-transcriptional regulation events from long-read sequencing data in *Arabidopsis thaliana*. In particular, we focus on the maturation of two chloroplast genes, *ndhB* and *ndhD*. These datasets provide rich ground for causal analysis: multiple post-transcriptional events, block-structured missing observations, and rare but biologically meaningful occurrences.

The reference work of Guilcher *et al.* [1] proposed a chronological ordering based on a deterministic principle: the more often a regulation event is observed alone, that is without others, the earlier it occured. We argue that this principle essentially assume that the sequence of regulation events is deterministic: a maturation event strictly depends on its predecessors: it is triggered by them and cannot occur otherwise. This does not align with the causal and probabilistic framework of Pearl, where rare events can still exert strong causal influence and probabilistically some events can still happen without their causal predecessors occurring. Consequently, our approach departs from frequency-based ordering and instead builds on Bayesian networks and Pearl's structural causal models.

Our method proceeds in three stages: 1. **Causal discovery**: learning a directed acyclic graph (DAG) over maturation events using HC, PC, LiNGAM, and NOTEARS, with a stability selection strategy for the $\ell_1$-regularization parameter in NOTEARS. 2. **Causal inference**: estimating causal effects via do-calculus and counterfactual reasoning. Missing data are handled via an EM algorithm that jointly imputes missing values and estimates the Bayesian network. 3. **Chronology construction**: translating estimated causal effects into maturation timeline models, yielding hypotheses about how perturbations (e.g., blocking a processing step) shift downstream events.

We show that this causal-chronology approach produces timelines that often differ substantially from the reference order in [1], and that it provides better fit, stability, and interpretability. Beyond improving chronology reconstruction, our framework allows "what if" questions and guides the design of targeted perturbation experiments in *Arabidopsis thaliana*.

# 2 Related work

## 2.1 Classical Methods for Gene Regulatory Network Inference

Early computational approaches to gene regulatory network (GRN) inference relied heavily on classical statistical and machine learning techniques designed to extract dependencies from large-scale gene expression data. A widely used family of methods is based on *information theory*, particularly mutual information (MI), which quantifies statistical dependencies between gene expression profiles. Algorithms such as **ARACNE** (Algorithm for the Reconstruction of Accurate Cellular Networks) apply MI combined with the data processing inequality to prune indirect associations, thereby yielding clearer networks [2]. Extensions like **CLR** (Context Likelihood of Relatedness) [3] and **MRNET** [4] further refine this principle by contextualizing MI scores or reducing redundancy among inferred interactions.

Another influential line of work exploits *tree-based ensemble methods*, most notably **GENIE3**, which decomposes the problem into regression tasks where each gene's expression is predicted from all others using Random Forests or Extra-Trees [5]. GENIE3 and its variants consistently rank highly in community benchmarks such as DREAM challenges, and they established machine learning as a reference framework for GRN inference. Complementary classical models include **Bayesian networks** [6], which encode conditional dependencies in a probabilistic graphical framework, and **structural equation models** (SEMs) [7], which leverage linear relations between variables while enforcing acyclicity constraints. These methods laid the groundwork for later developments in causal discovery and deep learning.

## 2.2 Temporal Data and Dynamic GRN Inference

While static methods infer regulatory interactions from steady-state or aggregated expression measurements, many biological processes unfold dynamically. To capture this temporal dimension, classical extensions of GRN inference integrate *time-series data*. For example, **dynGENIE3** extends the Random Forest-based GENIE3 by embedding regulatory effects in *ordinary differential equations* (ODEs), enabling simultaneous modeling of steady-state and temporal expression patterns [8]. Other approaches utilize *dynamic Bayesian networks* (DBNs) [9], which explicitly model temporal dependencies and can infer time-lagged interactions. Techniques such as **SINCERITIES** [10] infers directed and signed gene-regulatory edges by regressing each gene's expression on candidate regulators across time-stamped single-cell data, incorporating statistical distances (e.g. Kolmogorov–Smirnov) between distributions at successive time points to capture dynamics. More recently **GRNBoost2** [11] applies efficient gradient boosting regressors, where regulator–target interactions are ranked via feature importance in boosted decision trees.

These temporal frameworks are particularly relevant to the analysis of *full-length transcriptome datasets*, such as those generated by nanopore sequencing of plastid RNAs in *Arabidopsis thaliana* [12]. In this setting, the order and co-occurrence of post-transcriptional maturation events are not merely static associations but dynamic processes unfolding over time. This makes temporal GRN inference a crucial methodological bridge between raw sequencing data and interpretable models of chloroplast gene regulation.

### 2.3 Causality and GRN Inference

A central limitation of both static and temporal GRN inference methods is that they often rely on correlations or statistical dependencies, which do not necessarily imply causation. To support biological interpretation and experimental design, it is essential not only to infer structure, but also to reason about what would happen under perturbations. Causal discovery provides a principled framework that integrates graphical models, do-calculus, and structural equation modeling, thereby enabling the formal characterization of interventions and counterfactual reasoning (i.e. what would expression look like if we force a regulator on or off) [13]. For example, LiNGAM exploits non-Gaussianity to orient edges; PC and HC use conditional independence tests to recover causal structure; and NOTEARS [14] frames DAG learning as continuous optimization while preserving acyclicity, making it more amenable to counterfactual queries. These causal approaches provide a principled route beyond correlation-based GRNs—allowing not only inference of regulatory architecture, but simulation of interventions and estimation of causal effect—for instance, predicting how blocking or activating a transcript processing event may shift downstream maturation timelines, a perspective directly relevant in the full-length transcriptome context of *Arabidopsis thaliana*.

## 3 A Causal Perspective on Maturation Timelines

### 3.1 Dataset: Transcriptional Maturation Timelines in *Arabidopsis thaliana*

Our case study focuses on full-length transcriptome datasets describing the post-transcriptional maturation of two chloroplast genes in *Arabidopsis thaliana*, **ndhB** and **ndhD** [1]. Each dataset records for each sequencing read the presence or absence of specific maturation events, such as RNA editing or splicing, encoded as binary random variables ($1$ = observed, $0$ = not observed). Errors introduced by nanopore sequencing are treated as unobserved events and assigned the value $0$ during pre-processing. The *ndhB* dataset comprises 1 899 observations (or reads), of which only 117 are fully observed, and includes 12 distinct maturation events. Missing values are present in most reads and typically appear as contiguous blocks rather than isolated sites. In comparison, the *ndhD* dataset contains 7 752 observations with 930 fully observed, and involves 5 maturation events. Its missing values follow a similar block structure to those of *ndhB*. A glimpse of the untreated datasets is provided in Table 1.

The distribution of missing values makes it impractical to restrict the joint analyses of all events to fully observed reads, as this would drastically reduce the sample size (to 117 for *ndhB* and 930 for *ndhD*).
Instead, imputation strategies are required to exploit the full dataset. Furthermore, the block-structured nature of missingness suggests systematic sequencing coverage gaps rather than random dropout, which must be considered when designing statistical models.

These datasets provide a rich ground for causal discovery: the binary encoding of maturation events is well suited for Bayesian network learning, while the relatively high number of observations allows for robust estimation despite missing data.

Table 1: Untreated working datasets

(a) *ndhB*. The editing events are defined in terms of their genomic sites: ed1 = 94622, ed2 = 94999, ed3 = 95225, ed4 = 95608, ed5 = 95644, ed6 = 95650, ed7 = 96419, ed8 = 96439, ed9 = 96457, ed10 = 96579, ed11 = 96698, ed12 = 97016.

|      | ed1   | ed2   | ed3   | ed4  | ed5  | ed6  | ed7  | ed8   | ed9   | ed10  | ed11 | ed12 | intron |
|------|-------|-------|-------|------|------|------|------|-------|-------|-------|------|------|--------|
| 1582 | NaN   | NaN   | False | True | Err  | Err  | True | False | False | False | NaN  | NaN  | False  |
| 1095 | False | False | NaN   | NaN  | NaN  | NaN  | NaN  | NaN   | NaN   | NaN   | NaN  | NaN  | NaN    |
| 1495 | False | Err   | Err   | True | True | True | NaN  | NaN   | NaN   | NaN   | NaN  | NaN  | NaN    |
| 1060 | True  | False | True  | NaN  | NaN  | NaN  | NaN  | NaN   | NaN   | NaN   | NaN  | NaN  | NaN    |
| 830  | False | True  | True  | True | True | True | True | False | False | True  | True | True | True   |

(b) *ndhD*. The editing events are defined in terms of their genomic sites: ed1 = 116281, ed2 = 116290, ed3 = 116494, ed4 = 116785, ed5 = 117166.

|      | ed1   | ed2   | ed3   | ed4   | ed5   |
|------|-------|-------|-------|-------|-------|
| 4624 | True  | True  | True  | True  | True  |
| 171  | False | False | False | False | NaN   |
| 5637 | NaN   | NaN   | NaN   | NaN   | False |
| 4685 | True  | Err   | True  | True  | True  |
| 759  | False | False | True  | True  | True  |

### 3.2 From Frequency-Based to Causal Chronologies

The primary objective of this study is to infer a chronology of post-transcriptional maturation events for the *ndhB* and *ndhD* genes in *Arabidopsis thaliana*. A chronology here refers to a directed acyclic graph (DAG) where nodes represent maturation events and directed edges indicate temporal precedence (i.e. if there is an arrow from event $A$ to event $B$, then $A$ occurs before $B$ in the maturation timeline).

The postulate that guides the argument in [1] for proposing a chronology of maturation events is the following: *the more often an event occurs (or is observed) without others, the sooner it is the maturation timeline*.

Let us briefly summarize the pipeline elaborated in [1] to infer such a chronology starting from this assumption. According to the working dataset, the maturation events are viewed as random variables that can take binary values: $1$ if the event has been observed during the post-transcriptional maturation process, $0$ otherwise.

**Dependence discovery**. The first step consists in carrying out statistical pairwise dependence tests[1] in order to find out the actual dependencies between all possible couples of maturation events. In this step, we obtain an edge $A - B$ between two maturation events $A$ and $B$ if they are declared statisticaly dependent (adjusted p-value less than a pre-specified treshold). Note that by the end of this procedure an undirected graph is obtained.

**Temporal order decision**. Next, in order to decide whether regulation $A$ occured before or after regulation $B$ in the maturation timeline, one has to decide the direction of the edge by defining an arrow. For this, one applies the postulate:

$$\text{If } \#\{(A = 1, B = 0)\} > \#\{(A = 0, B = 1)\}, \text{ then } A \to B.$$
$$\text{If } \#\{(A = 1, B = 0)\} < \#\{(A = 0, B = 1)\}, \text{ then } A \leftarrow B.$$

Note that by the end of this procedure a directed graph without cycles (i.e. a *DAG*) is produced.

---
[1]*Exact Fisher tests* (ommiting missing data), to be more precise.

**Chronology representation**. Finally, we must account for two particular situations. First, the analysis does not always yield a complete chronological order: ambiguities may arise when determining the temporal relationship between two events. In such cases, we consider these events to occur simultaneously. Second, some maturation events occur independently of all others. These are either excluded from the chronology or represented in the DAG as isolated nodes.

By way of example, Figure 1 shows the complete analysis workflow for *ndhD* obtained in [1]. Figure 1a displays the dependencies network between different maturation events, while Figure 1b presents the final maturation timeline in its DAG form.

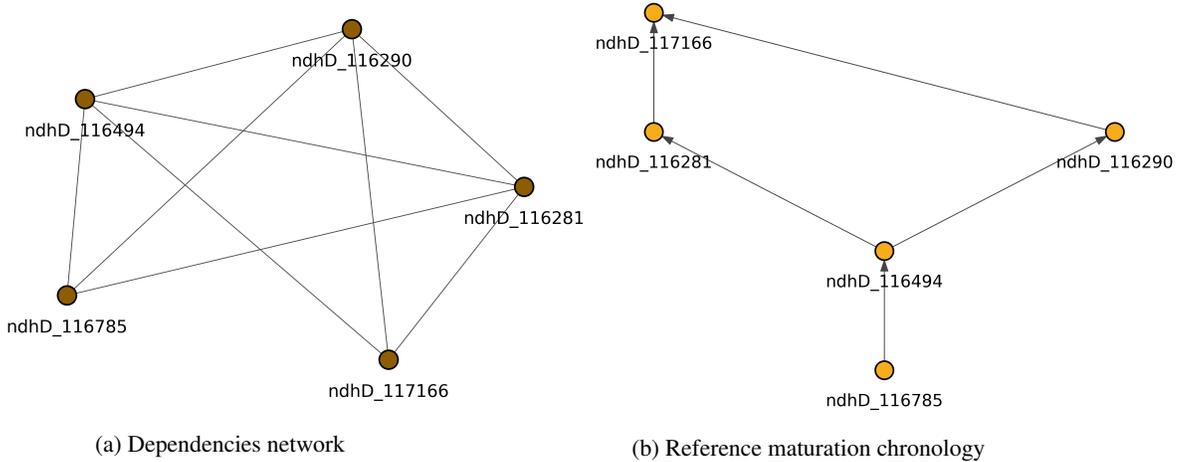

(a) Dependencies network

(b) Reference maturation chronology

Figure 1: Analysis of genetic regulation for *ndhD* from [1]: (a) statistical dependencies between maturation events, and (b) resulting directed acyclic graph representing the maturation timeline based on frequency ordering.

The previous conception is based on a determinist viewpoint, commonly shared in the biological community: *a maturation event strictly depends on its predecessors: it is triggered by them and cannot occur otherwise*. However, we know by now that the interactions between different maturation events can be much more complex and intricate than that (cf. [15], [16]. [17], [18], [19]).

In this sense, our approach represents a paradigm shift based on the following two objections to the deterministic postulate: *an event can have a strong causal effect even if it is rare* and *probabilistically some events can still happen without their causal predecessors occurring*. Accordingly, a maturation event can appear on the top of the maturation timeline even if it is hardly observed alone (that is without other regulations).

A biological explanation for this phenomenon relates to the speed at which events occur. Assume for example, that event $A$ strongly or rapidly promotes event $B$, then $B$ will quickly follow $A$, so $A$ will rarely be observed alone. If we further assume that $B$ can occur (for some small probability) without $A$ and $B$ does not promote $A$, then $B$ will often be observed without $A$. In this scenario $A$ is the cause of $B$, yet using the deterministic viewpoint we would conclude otherwise.

Such circumstances can indeed be observed within the working dataset. Namely, generating the frequency table of all five possible outcomes for *ndhD*, one can find instances of this phenomenon. Specifically, Table 2 shows the contingency matrix for the pair of maturation events located at the genomic sites 116494 and 116785. Let $A :=$ ndhD_116494 and $B :=$ ndhD_116785. We observe that ndhD_116494 appears much less frequently alone than ndhD_116785. According to the deterministic postulate, this would suggest an arrow ndhD_116785 $\longrightarrow$ ndhD_116494, as shown in Figure 1b. However, the causal Hill-Climbing model (HC-model, [20]), for example, proposes the reverse direction, ndhD_116494 $\longrightarrow$ ndhD_116785, as seen in Figure 2[2].

---

[2] In fact, most of our causal models propose this arrow (see Figure 10).

Table 2: Contingency matrix for the pair (ndhD_116494, ndhD_116785)

|                   | ndhD_116785 = 0 | ndhD_116785 = 1 |
|-------------------|-----------------|-----------------|
| ndhD_116494 = 0   | 82              | 144             |
| ndhD_116494 = 1   | 39              | 304             |

As we will see in Section 4, some of the proposed models even yield a reversed chronology compared to the one proposed in [1]. The causal HC-model again serves as an example: the path ndhD_116290 $\longrightarrow$ ndhD_116494 $\longrightarrow$ ndhD_116785 from Figure 2 reflects a reversed timeline relative to that proposed by the reference model in Figure 1b.

Moreover, the frequency table allows the following counts:

- ndhD_116290 appears 8 times alone.
- ndhD_116290 appears 26 times together with ndhD_116494.
- ndhD_116290 appears 262 times together with both ndhD_116494 and ndhD_116785.

In other words, the event ndhD_116290 strongly promotes ndhD_116494, so it is rarely observed alone (and ndhD_116494 is often observed without ndhD_116290). Besides, we see an increasing number of co-occurrences with the subsequent events of the proposed timeline.

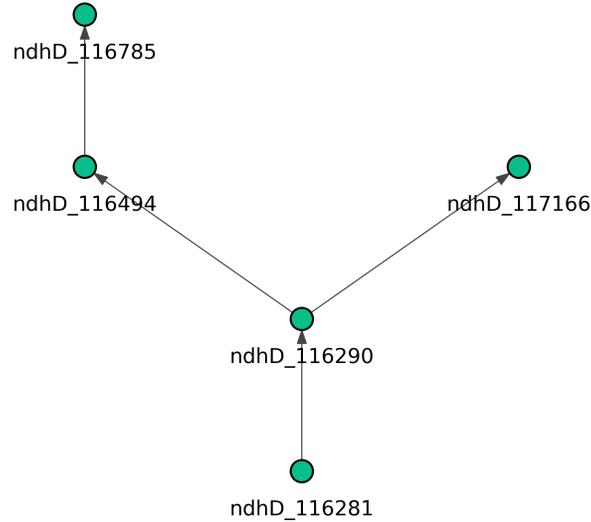

Figure 2: Causal HC-model for maturation chronology for *ndhD*

More formally, the deterministic postulate relies on a connection between the phenomenon $\left(A \text{ is the cause of } B\right)$ and the phenomenon $\left(A \text{ is observed alone more often than } B \text{ alone}\right)$. However, from a probabilistic point of view, it is clear that the second phenomenon does not allow to conclude any causal effect of $A$ on $B$. Therefore, our paradigm shift consists in adopting a probabilistic approach to the problem by directly computing $P(A \Rightarrow B)$. The mathematical framework for this is the causal inference theory in terms of *do-calculus* developed by J. Pearl (cf. [13], [21]).

**Definition 3.1.** We say that $A$ has a causal effect on $B$ if

$$P(B = 1 \mid \operatorname{do}(A = 1)) > P(B = 1 \mid \operatorname{do}(A = 0)).$$

We put $P(A \Rightarrow B) := P(B = 1 \mid \operatorname{do}(A = 1))$ and $P(\overline{A} \Rightarrow B) := P(B = 1 \mid \operatorname{do}(A = 0))$.

*Remark* 3.1. It is apparent that there is a close relationship between *chronology* and *causality*. Indeed, if $B$ occurs before $A$, then $A$ is definitely not a cause of $B$. In other words, causality entails a temporal order, hence a chronology.

To conclude this section, we summarize the proposed pipeline within our probabilistic framework, which is implemented and illustrated in Section 4.

1. **Causal discovery**. The first step is to infer a Directed Acyclic Graph (DAG) directly from the working dataset. We apply several well-known methods: Hill Climbing (HC, [20]), Peter-Clark (PC, [22]), LiNGAM [23]; and more recent continuous optimization variants such as NOTEARS [14]. This step involves both unconditional (statistical) dependence tests and conditional dependence tests, in order to establish directed edges between pairs of events in accordance with the observed data [3].

2. **Causal inference**. The first step yields a directed edge between two events, say $A \to B$. This suggests that $A$ is a potential cause of $B$, though it may nevertheless happen that $B$ occurs without $A$. To decide whether $A$ is an actual cause of $B$, we perform a causal inference analysis: isolating the causal effect; identifying and adjusting for confounders; and, when possible, designing interventions or applying graphical criteria such as the *backdoor criterion* to ensure that any spurious associations are removed.

3. **Chronology representation**. Finally, we construct the *maximal causality path* based on the previous analysis. More precisely, among all causal effects, we retain only those with the highest impact, respecting the topological order of the initial DAG obtained in the first step. By the end of this procedure one obtains a tree (as a sub-DAG of the discovered DAG in the first step) representing the chronology between the maturation events[4].

## 4 Inferring Causal Maturation Timelines in Arabidopsis thaliana via Bayesian Gene-Regulation Models

In this section, we apply the full causal inference pipeline described earlier to the post-transcriptional maturation of two chloroplast genes in *Arabidopsis thaliana*: **ndhB** and **ndhD**. These genes provide an ideal testing ground for our methodology: they involve multiple sequential maturation events—such as splicing and editing—captured in binary form from full-length transcriptomic nanopore reads, and exhibit block-structured missing data. Our goal is to infer robust and interpretable causal chronologies of these events, going beyond mere co-occurrence or frequency-based assumptions. Starting from imputed datasets, we learn causal structures using HC, PC, LiNGAM, and NOTEARS, estimate causal effects using do-calculus, and construct timeline DAGs that reflect high-impact, directed causal relationships between maturation events. The NOTEARS algorithm was adapted via a two-level stability selection strategy to enhance robustness in the discovery of causal structures; see Appendix Section 6.1 for details.

### 4.1 Imputation through Bayesian network discovery

An important limitation of the datasets used to study gene regulation in *Arabidopsis thaliana* is the significant presence of missing values. The distribution of these missing entries makes it impractical to restrict the dataset to fully observed rows, as this would drastically reduce the number of usable observations. This issue also affects the first step of our pipeline–*Causal discovery*–. As a result, we must perform imputation jointly with the learning of the corresponding Bayesian network.

To handle the full dataset, our strategy for missing data involves two stages of imputation: 1. ***Initial imputation***, and 2. ***Graphical model imputation***. Implementation details and configuration settings are available in a GitHub repository.

---

[3]This initial DAG contains a complex network of *all* possible (directed) interdependencies between different events.

[4]This tree reflects, by construction, the *relevant* (directed) interdependence network between different events. In this sense, it represents the *actual* causal structure underlying the initial DAG according to the *causal minimality principle* (cf. [13]).

1. **Initial imputation**. First, we perform a preliminary imputaton of missing values in order to have a complete dataset as a starting point for learning the Bayesian Network (via the selected causal discovery algorithm). Technically, the initial imputation is carried out using the `IterativeImputer` class from the `scikit-learn` Python library (cf. [24]). This method models each feature with missing values as a function of other features, using those estimates for imputation in a round-robin fashion. At each step, one feature (i.e. column) is treated as the target and the others as predictors.

2. **Graphical model imputation**. Next, we apply an *Expectation-Maximization* (EM) algorithm to jointly refine the imputed values and the Bayesian Network. The EM algorithm alternates between two steps:

   - **E-step (Expectation):** Missing values are imputed using the current Bayesian Network. For each missing value:
     - The parents of the variable in the current DAG are identified.
     - The corresponding Conditional Probability Distribution (CPD) is used to compute the most likely value for the missing entry, given its parents' values.
     - The dataset is updated by replacing the missing value with this most probable estimate.
   - **M-step (Maximization):** A new Bayesian Network is learned from the updated dataset using the selected causal discovery algorithm (HC, PC, LiNGAM, or NOTEARS). This includes:
     - Re-learning the DAG structure from the completed data.
     - Re-estimating the CPDs.

These steps are repeated iteratively until convergence, defined as the point where the proportion of changed directed edges between iterations falls below 1%.

*Remark* 4.1. Numerical experiments with our gene regulation datasets have shown that the choice of initial imputation method does not significantly affect the resulting graphical model. For example, a simple *mode* imputation yields the same final network.

*Remark* 4.2. The **Graphical model imputation** method described above converges rapidly. In fact, in our experiments, the network stabilizes by the second iteration. One possible explanation for this efficiency lies in the way missing values are updated: when a variable with missing data is identified, the marginal probabilities required for its update are computed using the values obtained from the initial imputation. That is, even if some parent variables also had missing values originally, they are treated as observed, based on the initial imputation.

A more principled approach would involve applying a *Belief Propagation* algorithm (cf. [25], [26]) in step (iv.c) to account for the uncertainty in the imputed parent values. However, our experiments indicate that this variant is significantly slower and does not lead to any substantial refinement of the resulting Bayesian network. In practice, the final graph is effectively the same as the one obtained directly from the initial imputation, without applying any further iterative refinement steps. One potential reason for this behavior is that the *Belief Propagation*-based version of the imputation process imposes, by construction, rigid structural constraints on the DAG, which in turn limits the ability of the EM algorithm to refine the network structure.

In our case study, we focus on datasets that describe the genetic regulation of the *ndhB* and *ndhD* genes located in the chloroplast of *Arabidopsis thaliana*. We provide a brief description of the datasets used and the distribution of missing (`nan`) values.

Maturation events are modeled as binary random variables: 1 if observed during post-transcriptional maturation, 0 otherwise. Errors introduced by the nanopore sequencing technology used in [1] are treated as unobserved events and assigned a value of 0 during initial pre-processing—an approach consistent with the binary encoding.

*ndhB* **dataset**
- **No. observations**: 1899 including 117 fully observed.
- **No. maturation events**: 12 whose sites are 94622, 94999, 95225, 95608, 95644, 95650, 96419, 96439, 96457, 96579, 96698, 97016.
- **Distribution of `nan` values**. Each observation contains a *unique* block with consecutive

reads, i.e. there is no isolated reads in the genomic region between nan values. Figure 3 shows a visualisation of the distribution of the values of the *ndhB* dataset before the imputation process described previously.

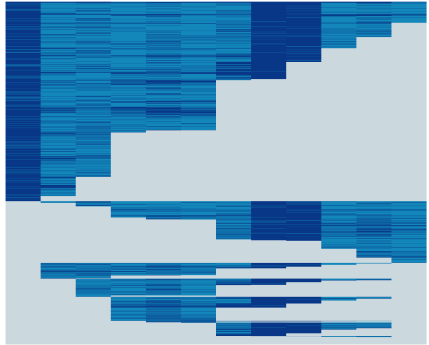

Figure 3: Distribution of values for *ndhB* before imputation: ■ = value 0, ■ = value 1, ■ = value nan

*ndhD* dataset
- **No. observations**: 7752 including 930 fully observed.
- **No. maturation events**: 5 whose sites are 116281, 116290, 116494, 116785, 117166.
- **Distribution of nan values**. Each observation contains a *unique* block with consecutive reads, i.e. there is no isolated reads in the genomic region between nan values. Figure 4 shows a visualisation of the distribution of the values of the *ndhD* dataset before the imputation process described previously.

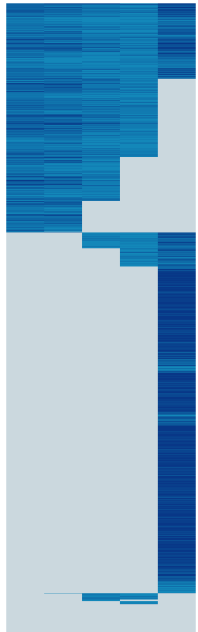

Figure 4: Distribution of values for *ndhD* before imputation: ■ = value 0, ■ = value 1, ■ = value nan

## 4.2 Causal inference

In this section, we detail the implementation of our probabilistic approach applied to the *ndhB* and *ndhD* genes of the chloroplast of *Arabidopsis thaliana*.

The first step in our pipeline is ***Causal discovery***. The graphs in Figure 5 and Figure 6 represent the discovered DAGs from the working datasets (i.e. after the imputation process described in Section 4.1), obtained using the HC, PC, LiNGAM, and NOTEARS algorithms. The HC and PC algorithms are implemented via the `pgmpy` Python package (cf. [27]), while LiNGAM is implemented using the `causal-learn` package (cf. [28]). Implementation details and configuration settings are available at our GitHub repository.

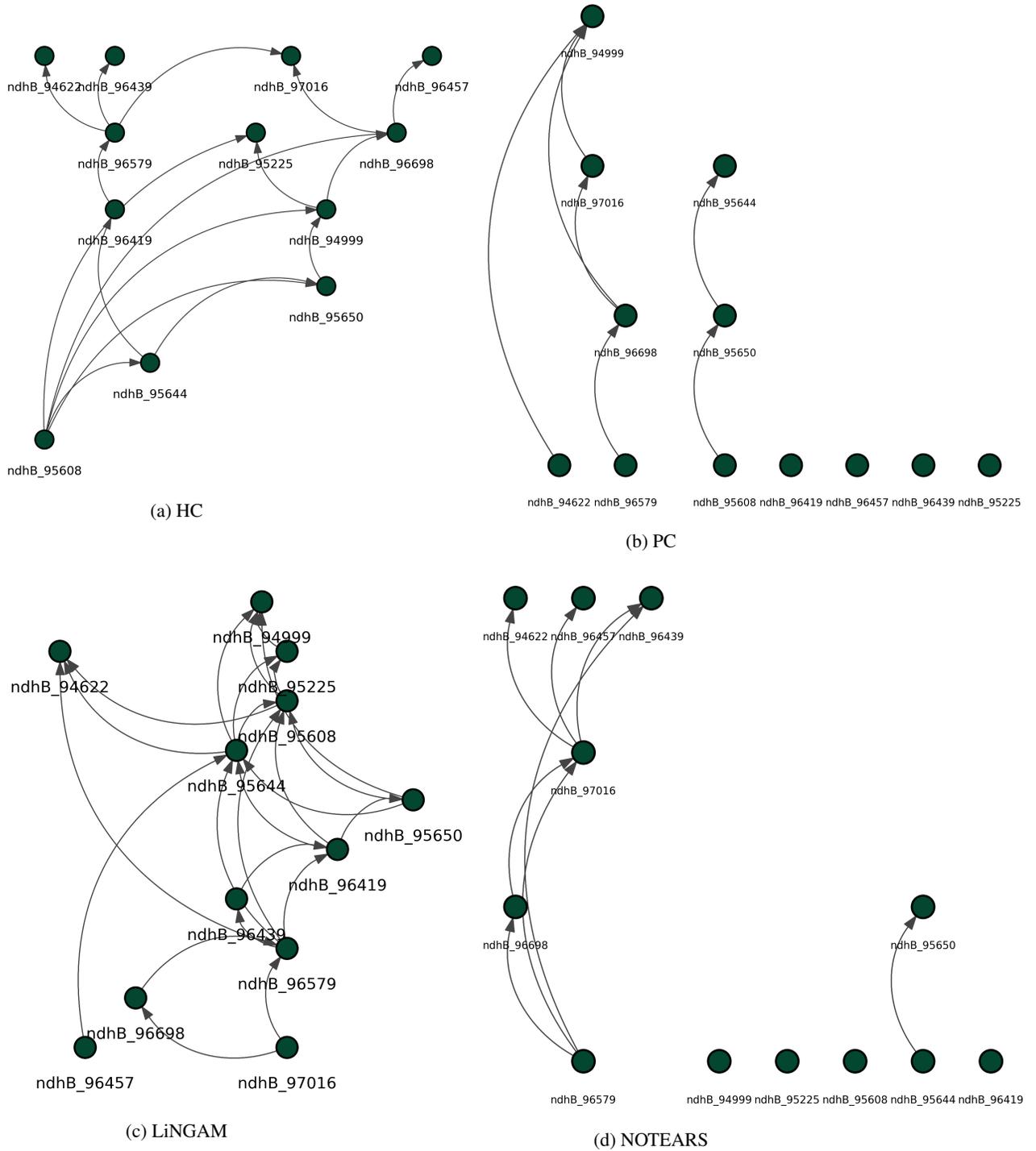

Figure 5: Causal Discovery for *ndhB*

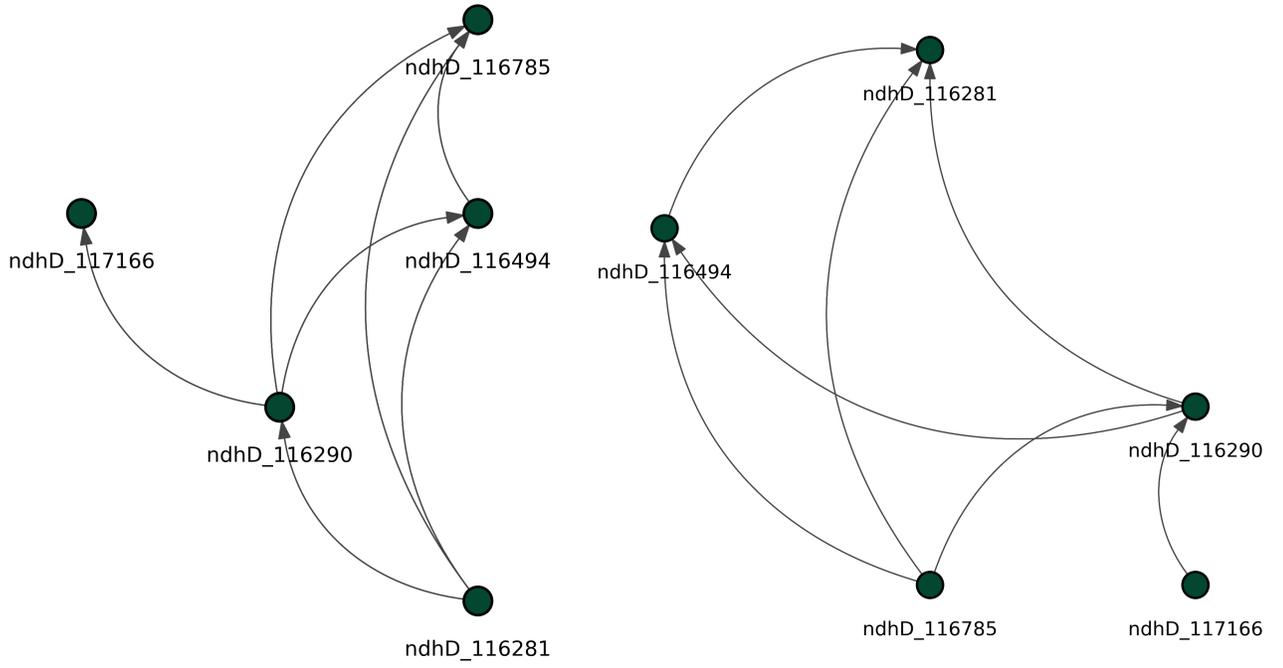

(a) HC

(b) PC

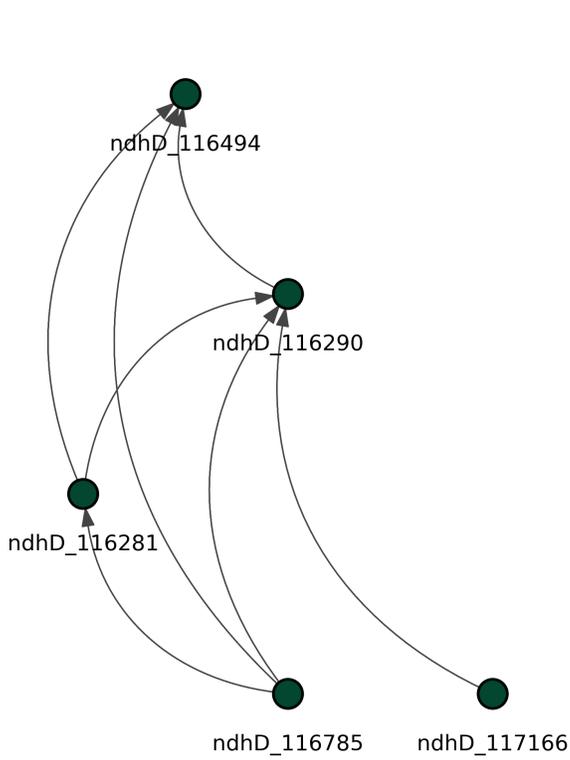

(c) LiNGAM

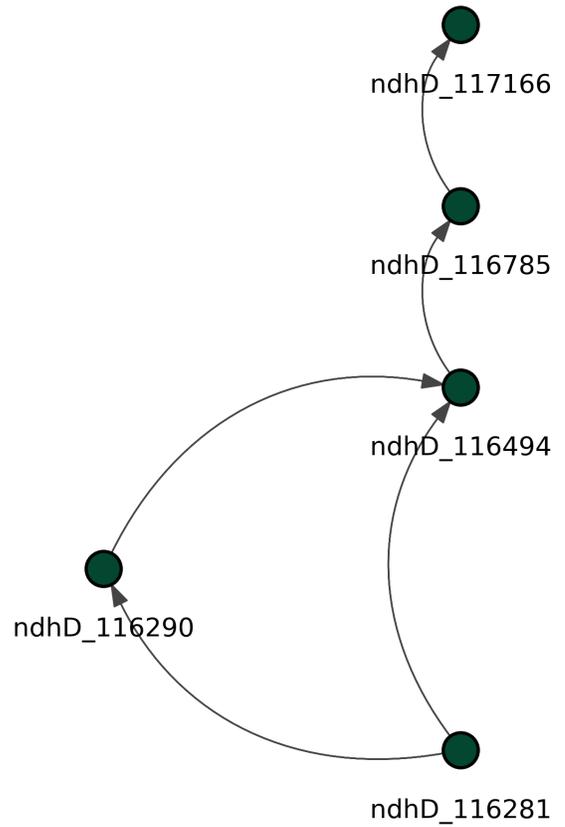

(d) NOTEARS

Figure 6: Causal Discovery for *ndhD*

#### 4.2.1 Interventional reasoning

Having identified directed edges through causal discovery, we now turn to the question of whether these edges reflect true causal relationships. To assess this, we apply **interventional reasoning** using the `DoWhy` Python library [29], which enables us to estimate the **Average Causal Effect (ACE)** of one event on another.

The **ACE** quantifies how much the outcome $Y$ would change on average if we were to intervene and set the treatment variable $X$ to 1 versus 0. Formally, it is defined as

$$\text{ACE} := \mathbb{E}[Y \mid \text{do}(X = 1)] - \mathbb{E}[Y \mid \text{do}(X = 0)].$$

In situations where a mediator lies along the causal path from $X$ to $Y$, we instead focus on estimating the Natural Direct Effect (NDE), which captures the component of the effect not transmitted through the mediator.

To carry out this analysis, we follow a structured four-step process for each DAG learned from the data:

1. **Model creation**. For each pair of events $(A, B)$ connected by a directed edge $A \to B$, we construct a `CausalModel` in `DoWhy`, designating $A$ as the treatment and $B$ as the outcome.

2. **Identification**. Using the structure of the DAG, `DoWhy` checks whether the causal effect is identifiable from observational data. This step relies on graphical criteria such as the **backdoor** and **frontdoor** conditions derived from do-calculus.

3. **Estimation**. If the effect is identifiable, we estimate the ACE using statistical techniques such as linear regression or matching. When mediators are detected, the estimation is refined to isolate the NDE.

4. **Validation**. Finally, we retain only those causal links for which the estimated ACE is strictly positive, indicating a likely genuine causal influence. When an NDE is used, this value is reported instead. In both cases, the result is recorded in our summary tables under the label *Actual Effect*.

Full estimation details are provided in the appendix Section 6.2, specifically in the *Causal relations figures* (Figure 14 and Figure 15), which present the most relevant causal relationships inferred from each of the previously obtained graphs together with the corresponding *Actual Effect* value. We refer to the available material at our GitHub repository for further precisions.

#### 4.2.2 Chronology representation

The final step in our pipeline is ***Chronology representation***: we construct the *maximal causality path* based on the previous analysis. Among all valid causal relationships estimated previously, we retain only those with the highest impact in accordance with the topological order of the discovered DAG. The construction of the causal-chronology DAG is carried out in several steps.

i. We sort the rows of the causal inference tables in descending order of *Actual Effect*.

ii. Among the possible causes for the same 'Outcome (Y)', we retain only the one with the greatest *Actual Effect*. This yields a collection of *high-impact* cause-effect arrows, which we refer to as *strong_causal_relations*.

iii. We compute the topological order of the discovered DAG and determine the level of each node. This is done using the `topological_sorting()` function from the `igraph` Python library (cf. [30]). The nodes are then grouped by their topological level.

iv. We sort the directed edges in *strong_causal_relations* in accordance with the topological levels of their source nodes.

v. We construct the corresponding DAG, thereby defining the chronology.

vi. We verify whether all roots of the initial DAG are included in the chronology. If not, they are added as isolated nodes (at topological level 0).

vii. We add all nodes that were already isolated in the initial DAG as well as those that can become isolated after this construction.

*Remark* 4.3. Note that this construction implies that the resulting causal-chronology DAG is a (directed) *tree*. Indeed, in this procedure each node has at most one incoming edge (since only the strongest causal link is retained for each outcome) and cycles are prevented by the topological ordering.

The application of this method to the *ndhB* and *ndhD* genes of the chloroplast of *Arabidopsis thaliana* yields the maturation timelines represented in their DAG form in Figure 7 and Figure 8, respectively.

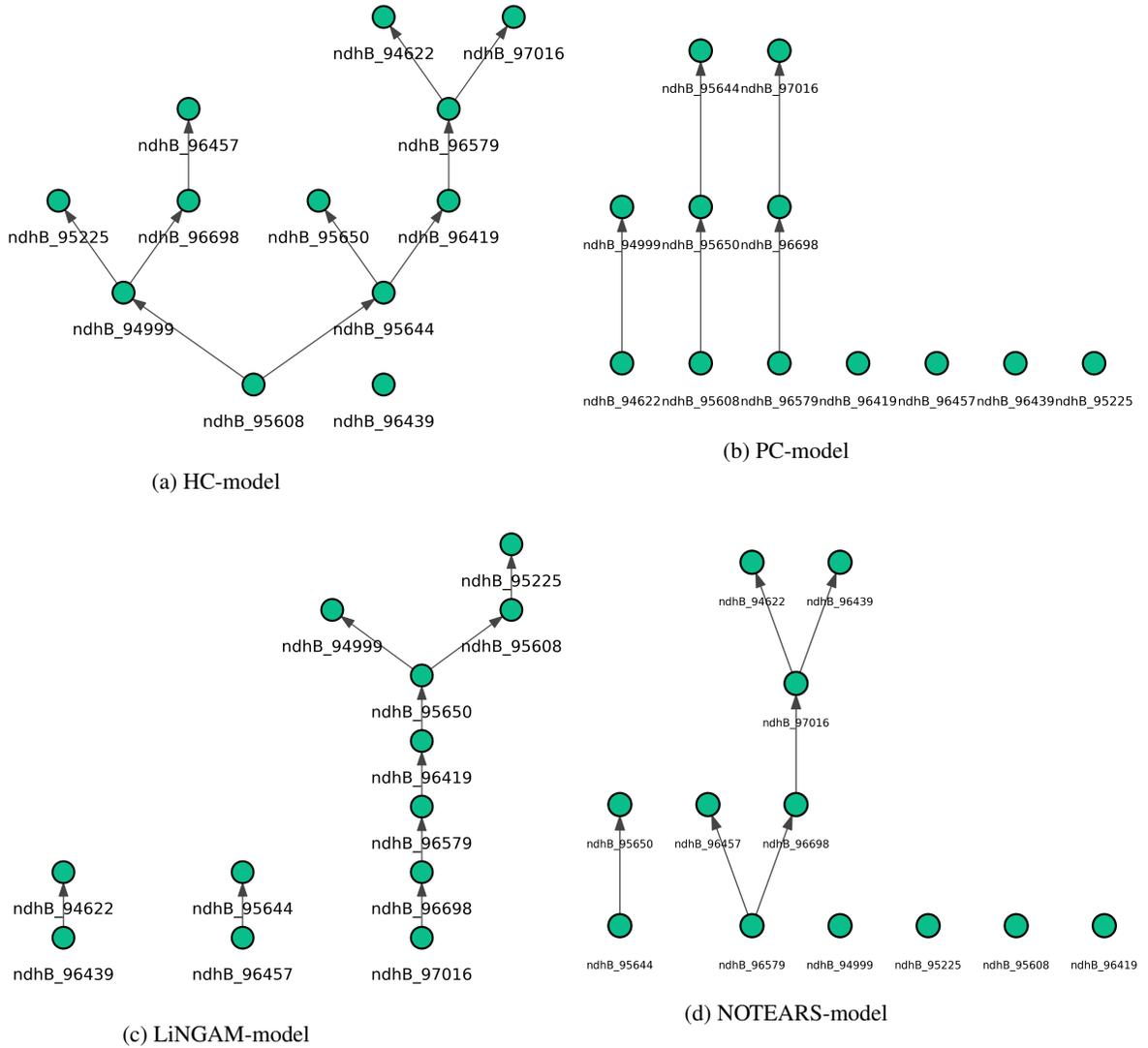

Figure 7: Causal models for maturation chronology for *ndhB*

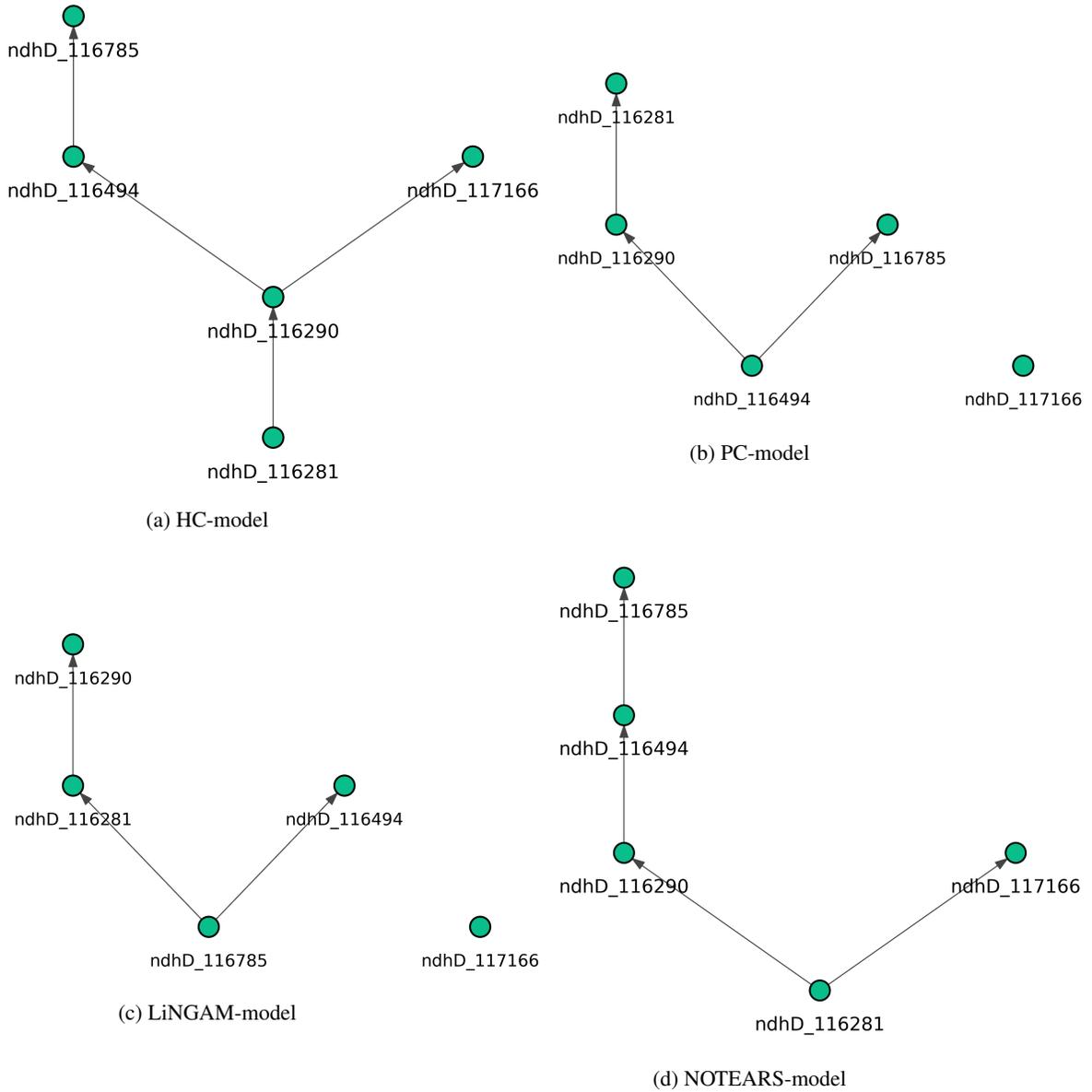

Figure 8: Causal models for maturation chronology for *ndhD*

These results naturally raise the question: *which model should be selected for a given gene?*

One approach is to validate each model against both the data and existing expert knowledge. Details of this validation process — including assessments of the causal estimators and the inferred DAG structures — are provided in Section 4.2.4.

Another complementary strategy is to perform a comparative analysis of the inferred networks across multiple models, aiming to identify **consistently recovered edges**. This consensus-based approach highlights stable relationships—whether directed or undirected—that appear robust to modeling choices. These consensus edges, combined with expert knowledge, can then be used to define a **whitelist** of trusted connections, guiding the construction of a more accurate and interpretable regulatory network.

For example, the cross-analysis of the causal models for *ndhB* (Figure 7) yields the sets of consistently recovered directed and undirected edges represented in Figure 9 as a multigraph.

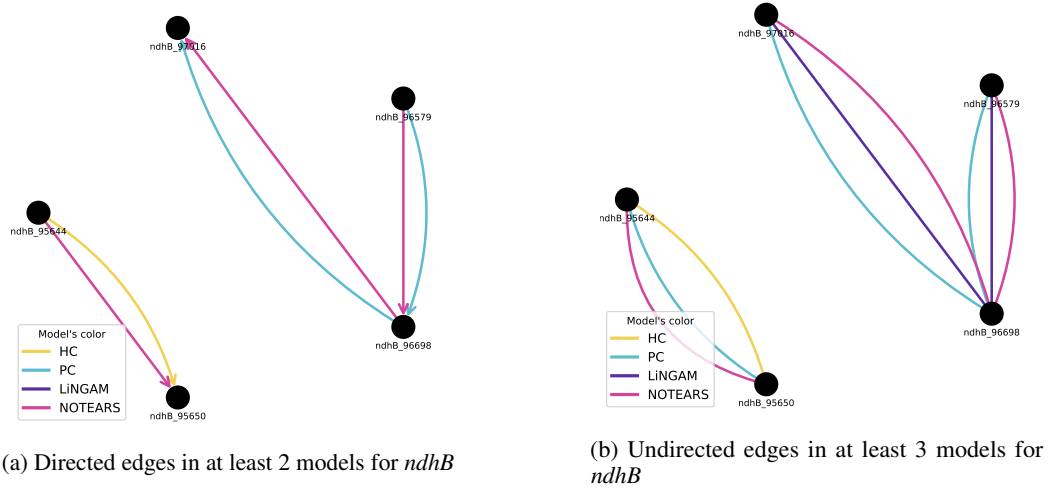

(a) Directed edges in at least 2 models for *ndhB*

(b) Undirected edges in at least 3 models for *ndhB*

Figure 9: Cross-analysis for consistently recovered edges within *ndhB* causal models

The cross-analysis of the causal models for *ndhD* (Figure 8) yields the sets of consistently recovered directed and undirected edges represented in Figure 10 as a multigraph.

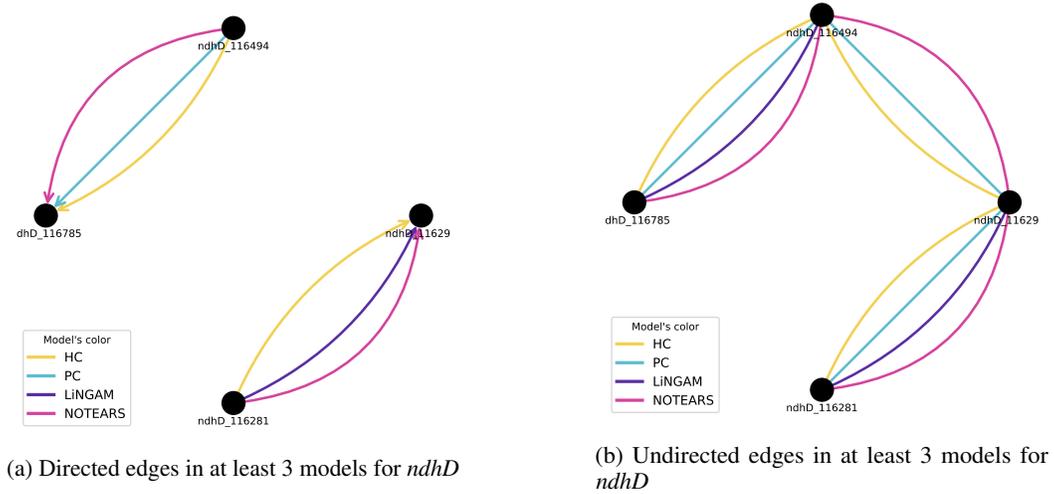

(a) Directed edges in at least 3 models for *ndhD*

(b) Undirected edges in at least 3 models for *ndhD*

Figure 10: Cross-analysis for consistently recovered edges within *ndhD* causal models

### 4.2.3 Comparison with Reference Models

We next compare our inferred maturation timelines with the reference chronologies presented by Guilcher *et al.* [1] to assess whether the causal-chronology models provide a superior fit to the observed maturation process.

The reference timelines for *ndhB* and *ndhD*, as defined in [1], are illustrated in DAG form in Figure 11.

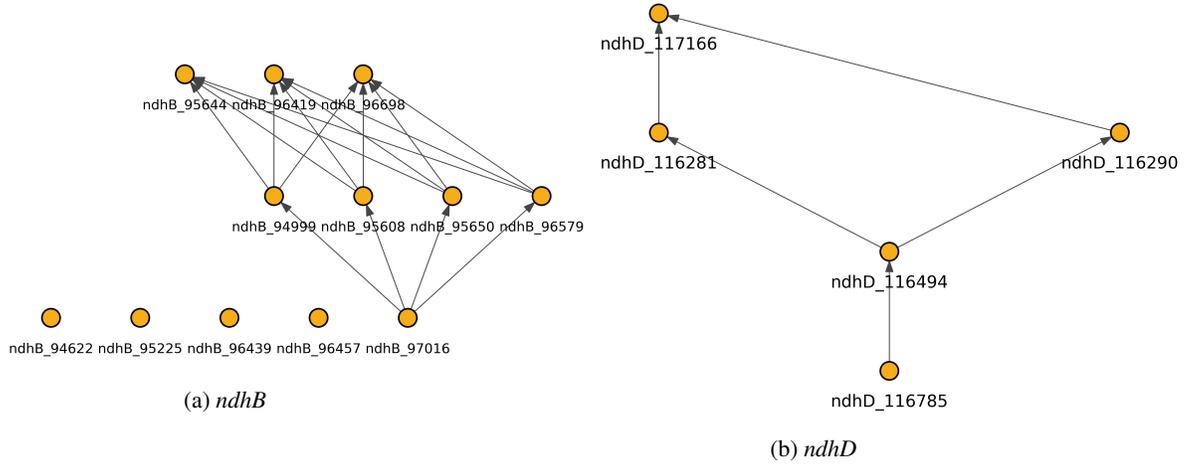

Figure 11: Reference models for maturation chronology

*Remark* 4.4. Note that the reference timeline for *ndhB*, illustrated in Figure 11, does not include the *intron*. This is because we observed that this event has an abnormally large effect on the others. In fact, all graphs we recovered with the deterministic or causal approach were unfalsifiable; removing this event makes them falsifiable. We decided to remove from the analysis which is further justified because, according to [1], the intron event occurred at the very end of the timeline.

To conduct the comparison, we evaluated both sets of models (causal-chronology vs. reference) using two standard metrics: the Bayesian Information Criterion (BIC) and the Log-Likelihood. The BIC balances model fit with complexity, penalizing overfitting, while the Log-Likelihood measures how well the model explains the observed data.

First, for each model we constructed a `DiscreteBayesianNetwork` using the `pgmpy` library ([27]), and estimated its Conditional Probability Distributions via a `BayesianEstimator`. We then computed `structure_score` (BIC) and `log_likelihood_score` on both the reference and causal-chronology networks. Implementation details and configuration settings are available in our GitHub repository.

The results, compiled in the charts of Figure 12, show that at least one causal-chronology model dominates the reference model in both BIC and log-likelihood. Importantly, *ndhB* shows three exceptions: the HC, PC and LiNGAM causal models underperform compared to the reference. Note however that the PC model is close to the reference model. Furthermore, performance differences are observed across causal discovery algorithms. These exceptions and performance variations are mainly due to the dataset size in terms of the number of variables. For instance, NOTEARS achieved better scores than the other methods for *ndhB*, but not for *ndhD*. Conversely, LiNGAM achieved better scores than the other methods for *ndhD*, but not for *ndhB*.

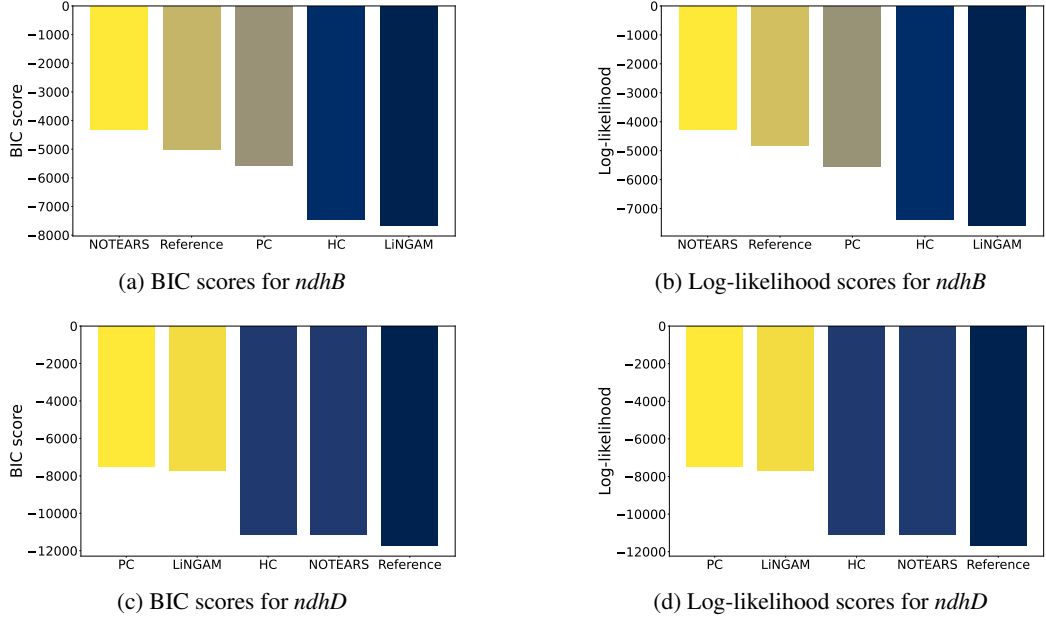

Figure 12: Bar plots of standard metrics for causal-chronology models and reference models: *higher is better*. Note that the scores (y-axes) are *negative values*.

A complementary metric used to compare our causal-chronology models to the reference is based on the Bayesian network structure itself, as explained in Section 4.2.4.2. Figure 13 summarizes the results of the falsification tests conducted to assess whether the corresponding DAG is accepted or rejected based on its fit with the provided data in terms of conditional independencies. Notably, we observe in Figure 13 that there are always at least two causal-chronology models that pass the tests with very high performance. As for the reference model, it is rejected for *ndhD* and accepted for *ndhB*.

#### 4.2.4 Reliability of the Proposed Methodology

To conclude our case study, we assess the **reliability** of the proposed methodology from two complementary perspectives: the stability of the estimated causal effects and the validity of the inferred graph structures.

**4.2.4.1 Robustness of causal effect estimates** The causal effects derived in our inference phase—such as the Average Causal Effect (ACE), Natural Direct Effect (NDE), and Natural Indirect Effect (NIE)—are influenced by modeling decisions, including the structure of the underlying graph and the statistical estimators used. To ensure that our results are not artifacts of specific configurations or assumptions, we subject these estimates to robustness checks.

These include tests that examine whether effect estimates remain stable under data perturbations, such as subsetting or introducing unrelated variables. Additional checks evaluate whether the estimates degrade appropriately when key assumptions are violated. Taken together, these tests help identify estimates that are likely reliable versus those that may be overly sensitive to modeling choices.

All reported causal effects (presented in Figure 14 and Figure 15) for which such analyses could be conducted passed the relevant robustness checks. Full results are available at our GitHub repository.

**4.2.4.2 Consistency of learned graph structures** Beyond individual effects, it is essential to validate the overall graphical models produced during causal discovery. A well-specified DAG should capture the key conditional independencies present in the data, and differ meaningfully from random alternatives.

To this end, we assess whether each graph implies conditional independence relations that are consistent with the observed data, and whether the graph lies in a narrow equivalence class—indicating a structure that is both specific and informative. These tests collectively provide insight into whether the learned DAG captures more than statistical noise or arbitrary correlations. The results (summarized in Figure 13) reveal that there are always at least two causal-chronology models that pass the tests with very high performance.

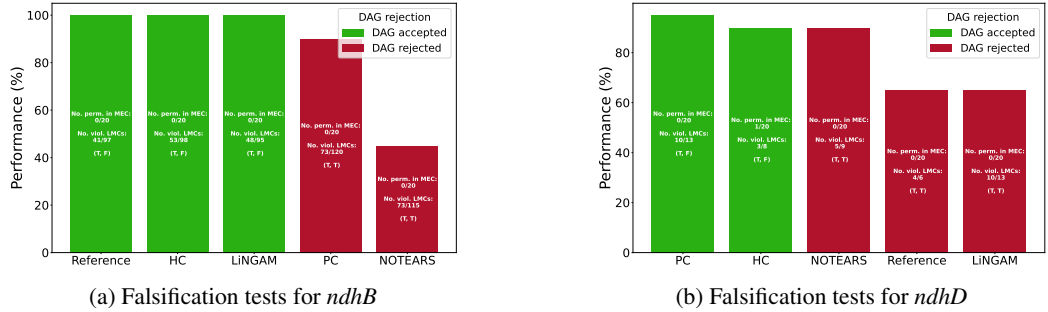

(a) Falsification tests for *ndhB*   (b) Falsification tests for *ndhD*

Figure 13: Falsification summaries

*Remark* 4.5. Among the four possible (*falsifiable*, *falsified*) outcomes:

- (T, F): the DAG is likely valid or at least not contradicted by the data.
- (T, T): the DAG is both specific and contradicted, therefore likely incorrect.
- (F, F): inconclusive; the structure may be underdetermined by the data.
- (F, T): an inconsistent outcome, suggesting possible issues with implementation or test assumptions.

It is worth noting that some causal structures—such as chains and forks—are probabilistically indistinguishable. Therefore, even when validation tests are passed, expert interpretation remains essential to resolve ambiguities and confirm causal plausibility.

For all these reliability assessments, we leverage the tools available in the `DoWhy` library.

## 5 Conclusive discussion

We have introduced a new methodology for reconstructing the chronology of genetic regulation events, grounded in Pearl's causal inference framework. This probabilistic approach represents a conceptual shift from earlier work in the field [1], moving beyond a deterministic viewpoint modeling to explicit reasoning over interventions. Our method captures the complex interplay of maturation events—consistent with prior findings in the specialized literature (e.g., [15], [16], [17], [18], [19])—and provides a well-founded strategy for extracting plausible maturation timelines from transcriptomic datasets.

The methodology proceeds in three stages: 1. Causal discovery (learning the DAG structure from data), 2. Causal inference (estimating the effects of interventions), and 3. Chronology construction (building a causal timeline as a tree-like graphical model).

We applied this pipeline to the post-transcriptional regulation of two chloroplast genes, *ndhB* and *ndhD*, in *Arabidopsis thaliana*, generating four alternative causal-chronology models for each gene. Each model corresponds to a distinct causal discovery algorithm. Compared to the reference chronologies [1], our causal models systematically achieved better fit for *ndhD*, with NOTEARS standing out for *ndhB*, according to both the Bayesian Information Criterion (BIC) and the log-likelihood. Furthermore, their robustness was assessed through multiple refutation tests—both on individual causal effect estimates and on the overall DAG structure—demonstrating a consistent advantage over the reference models for *ndhD*, while the reference model remains valid for *ndhB*.

We acknowledge that the causal discovery step introduces an element of uncertainty, particularly in the presence of latent confounders or limited data. As discussed in Remark 4.5, causal structure learning cannot be fully validated from observational data alone. To mitigate this limitation, we complemented standard approaches (HC, PC, LiNGAM) with a continuous optimization-based method: NOTEARS [14]. Because NOTEARS does not prescribe a principled way to set the $\ell_1$-regularization parameter—which governs the sparsity of the learned graph—we proposed a two-level stability selection procedure. This approach was benchmarked using the Sachs et al. dataset and shown to improve structural reliability. While [14] also reports results on this dataset, it provides little insight into parameter tuning or sensitivity.

A key practical challenge stemmed from the presence of numerous missing values in the dataset. Naively restricting the analysis to fully observed reads would have resulted in an unacceptable loss of statistical power. Likewise, truncating the variable space to ensure more complete rows would restrict biological interpretability and risk introducing structural biases. To overcome this, we adopted a two-step imputation strategy. First, we performed initial imputation using IterativeImputer from scikit-learn. Then, a custom graphical model-based EM algorithm iteratively refined the imputation and the DAG structure. Interestingly, the choice of initial imputation method proved to have minimal impact on the final graphical model.

Although our contribution is primarily computational, the proposed framework naturally yields testable hypotheses and experimentally verifiable predictions. The estimated causal effects are analogous to those obtained in randomized controlled trials and can therefore inform laboratory experiment design in a principled way. This opens the door to more sophisticated causal analyses, including counterfactual reasoning and targeted *what if?* questions, thereby enriching the interpretative power of long-read transcriptomic studies. These perspectives are aligned with discussions held with the project leaders of [1].

In summary, we present a principled, flexible, and data-driven framework for exploring gene regulation chronology. By combining causal discovery, statistical inference, and domain expertise, it establishes a foundation for reproducible, hypothesis-driven research in molecular biology.

# 6 Appendices

## 6.1 NOTEARS as a causal discovery algorithm

We adapted the NOTEARS algorithm [14] as part of our causal discovery pipeline. NOTEARS formulates the search for a DAG structure as a continuous optimization problem over a weighted adjacency matrix $W$, avoiding the combinatorial complexity of traditional structure learning. The acyclicity constraint is enforced using a smooth algebraic function, and sparsity is promoted via $\ell_1$ regularization. The result is a differentiable objective that can be optimized using gradient-based methods.

In our setting, we observed that NOTEARS was highly sensitive to the choice of the regularization parameter $\lambda$, with small changes leading to radically different graph structures. To address this, we introduced a two-level stability selection strategy (full details are available at our GitHub repository):

1. For a grid of $\lambda$ values, we performed multiple resampling iterations (bootstrap or subsampling) of the input data.
2. For each configuration, we recorded the set of directed edges returned by NOTEARS.
3. We defined a directed edge as *stable* if it appeared frequently across both resamplings and a contiguous range of regularization parameters, avoiding overly sparse or overly dense configurations.

This approach allowed us to construct a robust DAG by retaining only stable directed edges. It mitigates two known limitations of NOTEARS: its sensitivity to hyperparameters and the lack of invariance to variable rescaling [31].

While the resulting graphs were more interpretable and robust, it is worth noting that NOTEARS relies on a linear SEM assumption, and its optimization is non-convex. Therefore, learned graphs should be interpreted as hypotheses supported by data, rather than definitive causal claims. As such, we cross-validated NOTEARS outputs with those of HC, PC, and LiNGAM, and report only consistently supported edges in the final causal timelines.

## 6.2 Causal inference results

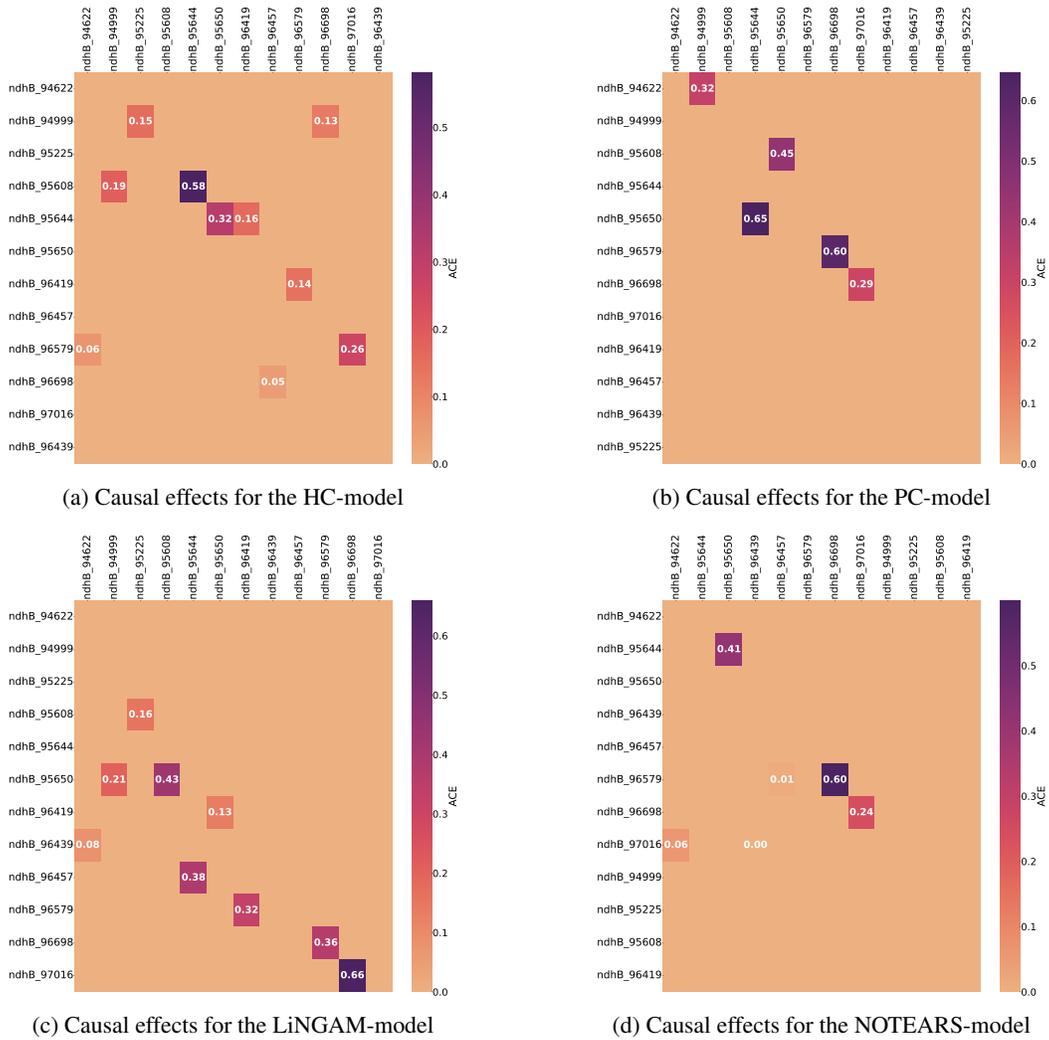

(a) Causal effects for the HC-model

(b) Causal effects for the PC-model

(c) Causal effects for the LiNGAM-model

(d) Causal effects for the NOTEARS-model

Figure 14: Causal relations for *ndhB*

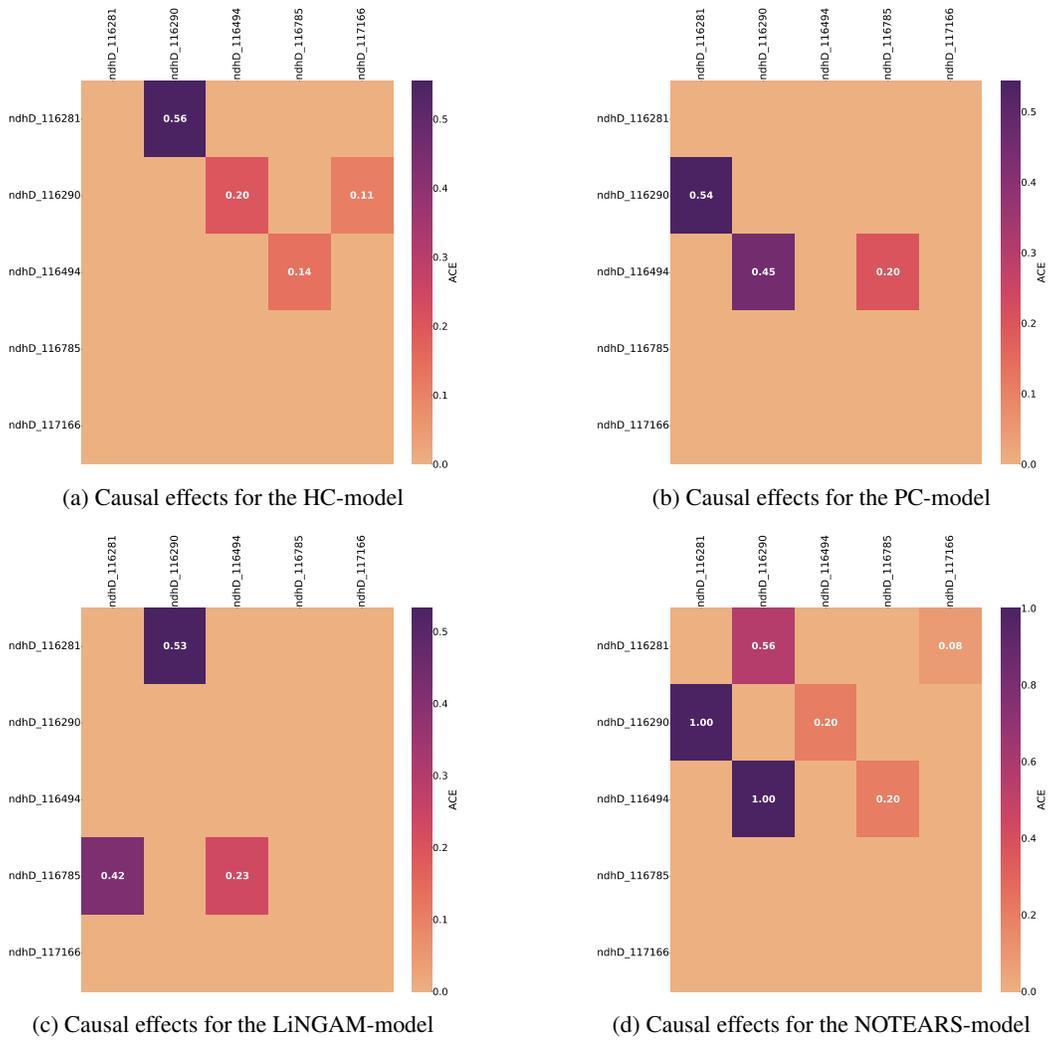

Figure 15: Causal relations for *ndhD*